\documentstyle[prl,aps,epsf,multicol]{revtex}
\begin{document}
\draft
\tighten

\title{Continuous 3d Freezing Transition in Layered Superconductors}
\author{Leon Balents\ $^1$ and Leo Radzihovsky\ $^2$}
\address{Institute for Theoretical Physics, University of California,
Santa Barbara, CA 93106-4030\ $^{1,2}$}
\address{Department of Physics, University of Colorado, Boulder,
CO 80309\ $^2$}
\date{\today}
\maketitle
\begin{abstract}
We use Ginzburg-Landau theory to study the $H_{c2}$ transition in
layered superconductors with field parallel to the layers, finding a
continuous 3d freezing transition to a triangular vortex super-solid
in the three-dimensional XY universality class. If screening effects
are neglected, off--diagonal--long--range--order survives only for
$d>d_{lc}=5/2$.  The partial breaking of the lowest Landau level
degeneracy induced by layering leads to a {\sl local} selection of a
triangular lattice structure, in contrast to the {\sl global} free
energy minimization in, e.g. Abrikosov's calculation.  Our results are
relevant to artificially layered superconductors and to strongly
anisotropic high T$_c$ materials.
\end{abstract}

\begin{multicols}{2}
\narrowtext

The development and application of the renormalization group in the
1970's and early 1980's greatly improved the understanding of
phase transitions.  Despite many dramatic successes resulting from
this creative explosion, the normal to superconducting transition
in finite field (NSFF), one of the earliest formulated examples of critical
behavior, is still not understood.

Any description of the NSFF transition begins with the
Ginzburg--Landau (GL)
free energy
\begin{eqnarray}
&&F[\psi,\vec A]=\int \! d^d{\bf r}
\bigg[|(\bbox{\nabla}+i{2\pi\over\phi_o}{\bf A})\Psi|^2\nonumber\\
&+& r_0|\Psi|^2 +{1\over2} g_0 |\Psi|^4+
{1\over8\pi\mu_0}|\bbox{\nabla}\times\bbox{A}-\bbox{H}|^2\bigg]\;,
\label{GL}
\end{eqnarray}
where $r_0\sim(T-T_c)/T_c$, $\phi_0=h c/2e$ is the flux quantum,
$\mu_0$ is magnetic permeability of the normal metal and is close to
unity, and we have for simplicity ignored the technically unimportant
effective mass anisotropy in the first term.  Eq.\ref{GL}\ respects
several symmetries which are broken within mean--field theory in the
mixed phase: translation and rotational invariance, as well as the
$U(1)$ gauge symmetry corresponding to off--diagonal
long--range--order (ODLRO) and the choice of phase of the pair
wavefunction $\Psi$.  Probably the most consistent definition of a
`gauge--invariant phase' was suggested by Moore\cite{Moore}, and is
completely specified by the positions of {\sl vortices}, or zeros of
$\Psi$.  A hidden third symmetry was introduced in
Ref.\onlinecite{FisherLee}\ as a (different!) $U(1)$ phase invariance
of the dual ``boson'' (dis-)order parameter $\psi$.  The expectation
value $\langle\psi\rangle \neq 0$ in the {\sl normal} phase.

The mean--field description of the NSFF transition was discovered by
Abrikosov\cite{Abrikosov}.  For fields smaller than $H = H_{c2}(T)$,
both $\Psi({\bf r})$ and $B$ become periodic, with an amplitude (of
$\Psi$ and of the periodic modulation of $B$) increasing continuously
from zero as $H$ or $T$ is reduced.  While a consistent (i.e. non-mean
field) theory of the phase transition is still lacking (but see
below), fluctuation effects are well understood within the lattice
phase. As discussed in Ref.~\onlinecite{Moore}, phonon fluctuations of
the vortex lattice destroy ODLRO for $d<4$.  However, the mixed phase
can persist with broken translational (and rotational) invariance.  In
principle, two types of lattice phases can exist: the ordinary solid,
with $\langle\psi\rangle=0$, and the supersolid, with
$\langle\psi\rangle \neq 0$\cite{FisherLee,FNF}.

The fluctuation corrected critical behavior has been studied in two
ways.  The more phenomenological approach attempts to construct a
theory in terms of vortices, the low energy degrees of freedom of the
mixed phase.  In this picture, the NS transition is accomplished by a
first order freezing of vortex lines from a liquid (the normal phase)
into a solid (the vortex lattice)\cite{NelsonSeung,Tesanovic}.
Somewhat more microscopic treatments begin with the Ginzburg--Landau
theory and attempt to capture fluctuation effects near the transition
directly in terms of $\Psi$.  As first pointed out in
Ref.\onlinecite{BNT}, the infinite degeneracy of the lowest Landau
level (LLL) requires a {\sl functional} renormalization group
treatment of the critical behavior in such models in high
dimensions\cite{LR,NewmanMoore}.  Unfortunately, the full connection
between this approach and the more physical picture of vortex lattice
melting remains unclear.

In this letter, we consider the case of a {\sl layered} superconductor
with magnetic field oriented parallel to the layers.  Because of the
partial breaking of LLL degeneracy imposed by this intrinsic
periodicity, we are able to carry the microscopic treatment much
further than in previous fluctuation studies.  This allows a direct
connection to be made between the Ginzburg--Landau and vortex
theories.  It also provides an analytic understanding of the selection
of a triangular vortex lattice: in all the previous studies to date
the breaking of degeneracy and ordering into a particular lattice has
only been captured by an explicit {\sl numerical} comparison of free
energies.

Our most important result is the demonstration of a {\sl
continuous} mechanism for three dimensional vortex freezing
in layered superconductors in parallel fields.

The effect of the CuO$_2$ layers (taken normal to the $y$ axis) can be
modeled using Eq.\ref{GL}\ and a spatially modulated local temperature
$r_0 = t_0 - \alpha\cos{(2\pi y/d)}$, where $d$ is the layer spacing.
For simplicity we first ignore screening (or equivalently consider
length scales below the London penetration length $\lambda$). With the
magnetic field $\bbox{B}=B\hat{\bbox{z}}$ along z-axis (or more
generally transverse to the xy plane, in $d=d_\perp+2$-dimensions), it
is convenient to work in the Landau gauge with $\bbox{A} = B
y\hat{\bbox{x}}$.  The quadratic part of the free energy is simplified
by expanding $\psi$ in the eigenfunctions of the corresponding
equation of motion. In this gauge, the eigenmodes are of the form
$\Psi = e^{i k_x x + i {\bf k_\perp\cdot r_\perp}}\phi$, with
\begin{eqnarray}
\bigg[- \!\partial_y^2 + (y-k_x\ell^2)^2/\ell^4\! +\! &\! r_o\! &\! -
\alpha\cos{(2\pi
y/d)}\bigg]\phi(k_x,\bbox{k}_\perp,y)\nonumber\\
&=&E(k_x,\bbox{k}_\perp)\phi(k_x,\bbox{k}_\perp,y)\;,
\end{eqnarray}
$E(k_x,{\bf k_\perp})=E(k_x)+k_\perp^2$ and $\ell=\sqrt{\phi_0/2\pi B}$.
For~$\alpha=0$ the
``Hamiltonian'' is diagonalized by Landau level (LL) eigenfunctions,
with the usual LL degeneracy of $E$ with respect to $k_x$.  The effect
of the periodic potential is to break the continuous translational
symmetry along $y$ down to a discrete group, which implies
$E(k_x+k_n,\bbox{k}_\perp)=E(k_x,\bbox{k}_\perp)$, where $k_n= d
n/\ell^2$ and $n$ is an arbitrary integer.  The original degeneracy
$N_\phi=L_x L_y/2\pi\ell^2$ is thereby reduced to $N_d=L_y/d$.  Each
LL is broadened into periodic bands with width proportional to
$\alpha$.  For our further approximations to hold, we consider a
strong field limit which consists of a number of constraints. First,
we take $\hbar 2e B/M c >> \tilde{\alpha}$ (see below), which implies
that LLs remain narrow on the scale of their separation, and allows a
clear separation of LLL physics. Secondly, we require the periodic
potential to vary smoothly on the scale of magnetic length,
i.e. $\alpha<<d^2/\ell^4$.  In this limit, the functions
$\phi(k_x,\bbox{k}_\perp,y)$ are Gaussian,
\begin{equation}
\Psi=\int_{k_x,\bbox{k}_\perp}\phi(k_x,\bbox{k}_\perp)
e^{i k_x x + i \bbox{k}_\perp\cdot \bbox{r}_\perp -
(y-k_x\ell^2)^2/2\ell^2}\;,
\label{phiDefn}
\end{equation}
where $\int_{k_x,\bbox{k}_\perp}\equiv\int d k_x
d^{d_\perp}\bbox{k}_\perp/(2\pi)^{(d_\perp+1)}$.
In the strong field limit the only effect of the periodic
potential is to introduce periodic $k_x$ dispersion, corresponding to the
energy cost of moving the center of the Gaussian along the cosine
potential. The quadratic part of the free energy becomes
\begin{equation}
F_0=\int_{k_x,\bbox{k}_\perp}\;
E(k_x,\bbox{k}_\perp)|\phi(k_x,\bbox{k}_\perp)|^2\;,
\end{equation}
where $E(k_x,\bbox{k}_\perp)=k_\perp^2 + t_0 +
\tilde{\alpha}\cos{(2\pi\ell^2 k_x/d)}$ and $\tilde{\alpha}=\alpha
e^{-(\pi\ell/d)^2}$.  It is convenient to work in the reduced zone
scheme where $k_x=k + d n/\ell^2$, and define new fields
$\phi_n(k,\vec{k}_\perp)\equiv\phi(k+ k_n,\vec{k}_\perp)$.
%
The nonintegral part $k \ll d/\ell^2$ describes small deviations
from the local minimum of the cosine potential.  Rewriting $F_0$ in
terms of the $\phi_n$ yields a $d_\perp\!+\!1$-dimensional
$N_d$-component~field~theory,
\begin{equation}
F_o=\sum_n^{N_d}\int_{k,\vec{k}_\perp}\;(k_\perp^2 +
\gamma k^2 + \tau)|\phi_n(k,\vec{k}_\perp)|^2\;,
\label{decoupledfreeenergy}
\end{equation}
where $\gamma=(\tilde{\alpha}/2)(2\pi\ell^2/d)^2$ and
$\tau=t_o+\ell^{-2}-\tilde{\alpha}$. The rotational invariance in
the $N_d$-dimensional $n$-isospin space of $F_o$ is of course broken
by the the full GL free energy, to which we now turn.

The quartic term is readily re-expressed in terms of the
$\phi_n(k,{\bf k_\perp})$ fields using Eq.\ref{phiDefn}. The result is
\begin{eqnarray}
&&F_{\rm int}=g\sum_{n_1\cdots n_4}\int_{k_1 \cdots k_4}\phi^*_{n1}(\bbox{k_1})
\phi_{n2}(\bbox{k_2})\phi^*_{n3}(\bbox{k_3})\phi_{n4}(\bbox{k_4})\nonumber\\
&&\times V\big[\{n_i\},\{\bbox{k_i}\}\big](2\pi)^{(d-1)}
\delta^{(d-1)}(\bbox{k_1}\!-\!\bbox{k_2}\!+\!\bbox{k_3}\!-\!\bbox{k_4}),
\label{Fint}
\end{eqnarray}
where
\begin{equation}
V= \left({1\over2\pi\ell^2}\right)^{1/2}
e^{-(d/\ell)^2\left[k^2+l^2\right]/4}
\delta_{n_1-n_2,n_4-n_3}.
\label{Vdef}
\end{equation}
Here $n_1=n/2+k/2, n_2=n/2+l/2, n_3=n/2-k/2,$ and $n_4=n/2-l/2$, and
we neglected small $O(k)$ corrections in the exponential in
Eq.\ref{Vdef}.

Infrared divergences emerging below $d_{uc}=5$ can be studied using
renormalization group (RG) methods in $d=5-\epsilon$ dimensions. We
use the momentum shell method, in which short wavelength modes in a
shell $\Lambda/b<p<\Lambda$ are progressively integrated out to yield
a renormalized free energy.  The RG is best formulated in terms of the
coupling matrix $u_{k,l}=V[n_1,n_2,n_3,n_4]$.  The matrix $u_{k,l}$
vanishes when $k,l$ are not simultaneously odd or even. The discrete
translational invariance along $y$ enforces the independence of $u_{k,l}$
on $n$.

We perform an infinitesimal RG transformation with $b=e^{dt}$, and
rescale $r \rightarrow b r$, $\phi_n \rightarrow b^{(3-d)/2}\phi_n$ to
obtain
\begin{equation}
\partial_t u_{k,l}=\epsilon u_{k,l}-2 m u_{k,s}\; u_{s,l} -
8 u_{s,s-k+l}\; u_{s-3k/2+l/2,s-k/2-l/2}
\label{rgflow}
\end{equation}
in the thermodynamic ($N_d \rightarrow \infty$) limit.  For later
purposes (see below) have considered an extended $U(1)\times O(m)$
theory with the generalized interaction $f_{int}\propto \phi^{*
i}_{n1}(\bbox{k_1}) \phi^j_{n2}(\bbox{k_2})\phi^{*
i}_{n3}(\bbox{k_3})\phi^j_{n4}(\bbox{k_4})$, with $i,j=1,\ldots,m$.

We have succeeded in finding only one exact (but unstable) solution to
the Eq.\ref{rgflow} with $u^*_{k l}=\epsilon/(2m+8) \delta_{k,0}
\delta_{k,l}$, corresponding to simultaneous but independent XY
transitions within each layer. For $m\rightarrow\infty$, the second
nonlinearity of Eq.\ref{rgflow} becomes negligible and the resulting
matrix equation is easily solved. The large $m$ fixed point solutions
are then any matrix with an arbitrary combination of $0$ and
$\epsilon/2m$ eigenvalues. The the most stable solution is
proportional to $\delta_{k,l}$, corresponding to an arbitrarily
long-range interaction $\propto\sum_{n,k} \phi^{*
i}_{n}\phi^j_{n}\phi^{* i}_{n+k}\phi^j_{n+k}$.  However, the form of
the initial condition for $u^o_{k,l}=u_k u_l $, Eq.\ref{Vdef},
dictated by the LLL physics, insures that the physical fixed point
solution is of the same short range form, but with a single nonzero
eigenvalue flowing to $\epsilon/2m$.  At this fixed point standard
methods give the correlation length exponent $\nu=1/(2-\epsilon/m)$,
which diverges (for $m=1$) in $d=3$, {\it naively} suggesting $3$ as
the lower critical dimension ($d_{lc}$). Although this result is in accord with
dimensional reduction arguments of $d\rightarrow d_\perp=d-1$, we will
show below that a more careful analysis leads to $d_{lc}=2.5$ for the
unscreened case, invalidating dimensional reduction. The breakdown of
dimensional reduction may also be important for the NSFF transition in an
isotropic superconductor.
Unfortunately, the $1/m$ corrections to the $m\rightarrow\infty$ fixed
points appear to be singular, and we have not succeeded in expanding
toward $m=1$.

Although we have not been able to solve above RG equations, if the quartic
interaction remains short-ranged (as is physically plausible from the
existence of the magnetic length scale $l$), much insight
into this freezing transition can be gained from the mft and from studying
the fluctuations in the low T lattice phase. The short-range form
of $V\big[\{n_i\},\{\bbox{k_i}\}\big]$
provides the key simplification relative to the problem without the layered
potential.

Eq.\ref{Fint} contains interactions between all of the
$\{\phi_n\}$ at all momenta.  We expect the minimal free energy to
occur for $k=0$, i.e. uniform $\phi_n$; higher order terms in a
gradient expansion are also less relevant in the RG
sense.
Because increasing $n_1-n_3$ and $n_2-n_4$ results in exponential
suppression, it is sufficient to keep only the largest few terms in the
interaction free energy
\begin{eqnarray}
F_{\rm int} & = & \int d^{d-1}{\bf r} \sum_n \bigg\{ {u \over
2}|\phi_n|^4 + {w\over 2}|\phi_n|^2|\phi_{n+1}|^2 \nonumber \\ & & + {v \over
2}(\phi_{n+1}^*\phi_{n-1}^*\phi_n^2 + c.c) \bigg\},
\label{fint}
\end{eqnarray}
where $u = 2g/\sqrt{2\pi\ell^2}$, $w =
4g/\sqrt{2\pi\ell^2}e^{-d^2/2\ell^2}$ and
$v=4g/\sqrt{2\pi\ell^2}e^{-d^2/\ell^2}$.  Since $w,v \ll u$, there
will be a broad crossover regime described by Eq.\ref{fint}\ with
$w=v=0$.  Specializing to $d=d_\perp+2=3$, Eqs.\ref{fint}\ and
\ref{decoupledfreeenergy}\ indicate that in this limit the individual
$\phi_n$ are decoupled and comprise {\sl two--dimensional} XY order
parameters in each layer.  These develop significant amplitude below
the mean--field transition at $\tau=0$, where the system shows
the usual broad specific heat maximum associated with the 2d XY
model. Physically, the large field limit leads
to a strong variation of the phase of $\psi$, destroying the Josephson
coupling between neighboring superconducting layers.

To study the behavior for $\tau <0$, we ignore (massive) amplitude
fluctuations and let $\phi_n = \sqrt{|\tau|/u} e^{i\tilde{\theta}_n}$.
Up to a constant, the free energy becomes
\begin{eqnarray}
F & = & \sum_n \int \! dx dy \bigg\{ K_x (\partial_x\tilde{\theta}_n)^2
+ K_z (\partial_z\tilde{\theta}_n)^2 \nonumber \\ & & +
\tilde{v}\cos(\tilde{\theta}_{n+1}+\tilde{\theta}_{n-1}-2\tilde{\theta}_n)
\bigg\},
\label{Fxy}
\end{eqnarray}
where $K_x = \gamma K_z = \gamma|\tau|/u$ and $\tilde{v} = \tau^2
v/u^2$.  For $\tilde{v}=0$, the system undergoes simultaneous
Kosterlitz--Thouless (KT) transitions within each layer (each $n$) to
decoupled quasi--long--range--ordered (QLRO) states for $K = \sqrt{K_x
K_y} > K_{\rm KT} \approx 4/\pi$.  Correlation between the phases in
the layers is established for $\tilde{v} > 0$.  A simple calculation
using standard methods for the 2d sine--Gordon model demonstrates that
$\tilde{v}$ is indeed a {\sl relevant} perturbation for any $K > K^* =
3/(4\pi) < K_{\rm KT}$, and therefore preempts the KT
transition\cite{JKKN}.  Instead, a true three--dimensional transition
occurs at a critical stiffness $K_c \lesssim K_{\rm KT}$ ($K_c$ in
fact approaches $K_{\rm KT}$ as $\tilde{v} \rightarrow 0$).

For $K > K_c$, the ordered state is determined by the minimization of
Eq.\ref{Fxy}.  Because $\tilde{v} >0$, the argument of the cosine
should be an odd multiple of $\pi$, and a convenient choice of
symmetry breaking is $\tilde{\theta}_{n,0} = 0, 0, \pi, \pi, 0, 0, \cdots$
on successive layers.  Examination of the zeros of $|\Psi|^2$,
Eq.\ref{phiDefn}\, shows that this low energy state is a triangular
vortex lattice with lattice
vectors $2\pi\ell^2 {\bf\hat{x}}/d$ and $\pi\ell^2 {\bf\hat{x}}/d+
d{\bf\hat{y}}$\cite{comment}.  The simple determination of the lattice
structure from the local $\tilde{v}$ phase interaction strongly contrasts
with the situation in the conventional unlayered GL theory, in which the
choice of lattice relies upon a complex non--local minimization in
the full LLL basis.

Fluctuations within the ordered phase are described by the phase
shifts $\theta_n \equiv \tilde{\theta}_n - \tilde{\theta}_{n,0}$,
which are governed by Eq.\ref{Fxy}\ with the sign of $\tilde{v}$
changed.  To study the long--wavelength properties of these phase
variations, the cosine may be replaced by a mass, leading to
\begin{equation}
F_{\rm eff} = \int \! d^3{\bf r} \bigg\{ \kappa_x (\partial_x\theta)^2
+ \kappa_z (\partial_z\theta)^2 + \kappa_y (\partial_y^2\theta)^2
\bigg\},
\label{Feff}
\end{equation}
where $\kappa_x = K_x/d$, $\kappa_z = K_z/d$, $\kappa_y = 2 m^2 d$,
and the thermally renormalized mass $m^2 = \tilde{v}^{8\pi K/(8\pi K -
6)}$.  A straightforward calculation shows that the phase fluctuations
$\langle |\theta|^2 \rangle$ are finite in three dimensions, with
$d_{lc}=2.5$, showing that GL theory (without screening) displays true ODLRO,
in contrast to the naive dimensional reduction conclusion of $d_{lc}=3$.
This free energy occurs in the study of liquid
crystals\cite{Toner} and is known as m=1, XY Lifshitz model (m=1, for one soft
$k^4$ component). The model can be rewritten as the zero field gauge theory
\begin{equation}
F_{\rm eff} = \int \! d^3{\bf r} \; \bigg\{\kappa_\perp
|\bbox{\nabla_\perp}\theta-\tilde{a}\bbox{y}|^2+
\kappa_y|\bbox{\nabla}\tilde{a}|^2\bigg\},
\label{Feff2nd}
\end{equation}
where $\tilde{a}\bbox{y}$ is a ``fake'' gauge field with only a
y-component and we have rescaled x and z coordinates.
Ref.\onlinecite{Toner}\ concludes that the approach to high
temperatures can proceed in two stages, via an intermediate
``orientationally ordered'' state.  This phase lacks ODLRO, but
retains order in $\partial_y\theta \sim u$, and hence retains broken
translational invariance.  It is also known\cite{Toner}\ that the free
energy, Eq.\ref{Feff2nd}, is self-dual, which implies that the
intermediate state has long-range correlations in the disorder
parameter $\psi$ -- i.e. it is a super-solid.  We see no a priori
reason why such a topologically ordered phase may not survive below
$d_{lc}$, e.g. in two dimensions.

Without screening, the solid to super-solid transition is therefore in
the 3d, $m=1$, XY Lifshitz universality class.  By analogy with the 3d
normal-to-superconducting zero-field transition in the strongly type
II limit\cite{DH}, we expect a {\sl continuous} transition.

What is the universality class of the normal to super-solid
transition?  As described in Appendix B of Ref.\onlinecite{smectic},
the only broken symmetry {\sl associated with the transition} is
translations along the $x$ axis.  A Landau formulation in terms of the
complex order parameter $\rho_{\bf\hat{x}} = \langle \Psi^\dagger\Psi
e^{i d x/\ell^2} \rangle$ implies the critical behavior is of the 3d
XY type (c.f. also Ref.\onlinecite{Toner}). The phase of
$\rho_{\bf\hat{x}}$ is proportional to the phonon coordinate $u$.
Note that electromagnetic screening does not play an important role
here, because long-range interactions are screened
anyway by the fluid of vacancies/interstitials  already present in the
super-solid.

The above considerations apply when
magnetic variations may be neglected.  Gauge field fluctuations can be
included (in the ordered phase) simply by the minimal coupling
prescription $\partial_\mu \theta \rightarrow \partial_\mu \theta -
a_\mu$, where ${\bf a} = 2\pi({\bf A} - By{\bf\hat{x}})/\phi_0$ and
$\phi_0 = hc/2e$ is the superconducting flux quantum.  The
gauge--invariant free energy is thus
\begin{eqnarray}
F_{\rm gauge} & = & \int \! d^3{\bf r} \bigg\{ \kappa_x
|\partial_x\theta - a_x|^2
+ \kappa_z |\partial_z\theta-a_z|^2 \nonumber \\
& & + \kappa_y|\partial_y(\partial_y\theta - a_y)|^2
+ {\phi_0^2 \over {32\pi^3}}|\bbox{\nabla}\times{\bf a}|^2
\bigg\}.
\label{gaugefreeenergy}
\end{eqnarray}
Note that, despite its unusual form, the $\kappa_y$ term is invariant
both under gauge transformations and the global shift
(c.f. Eq.\ref{Feff}) $\theta \rightarrow \theta + \lambda y$.
Eq.\ref{gaugefreeenergy}\ also encodes a {\sl transverse Meissner
effect}: minimizing $F_{\rm gauge}$ with respect to $\theta$ for ${\bf
a} = B_y z {\bf\hat{x}}$ gives a divergent free energy density for an
additional field along the $y$ axis.  Perturbing fields along $x$ and
$z$ are not screened.

We now choose a gauge with $\theta=0$ to determine the true
low--energy structure of the theory.  The result is a modification of
the usual Anderson--Higgs mechanism which occurs in the Meissner
phase: the $a_x, a_z$ components of the gauge field become massive,
but $a_y$ continues to describe a gapless mode.  This remarkable
mechanism, allowed by the unusual $\kappa_y$ term in
Eq.\ref{gaugefreeenergy}, has a direct physical interpretation.  Once
gauge fluctuations are included, only the {\sl zeros} of $\Psi$ are
gauge--invariant objects, and it is their displacements ${\bf u}$
which are the physical degrees of freedom.  Because of the potential
in the $y$ direction, only $u_x \equiv u$ is gapless.  It is
straightforward to derive the relation $u = - \ell^2 (\partial_y
\theta - a_y) \rightarrow \ell^2 a_y$ in the $\theta=0$ gauge.
Integrating out the massive $a_x$ and $a_z$ fields then gives the
elastic free energy
\begin{eqnarray}
F_{\rm el.} & = & {1 \over 2}\int \! d^3{\bf r} \; \bigg\{ c_\perp
|\bbox{\nabla}_\perp u|^2 + c_y |\partial_y u|^2 \nonumber \\ & & -
c_4 [\kappa_x^{-1}|\partial_x\partial_y u|^2 +
\kappa_z^{-1}|\partial_z\partial_y u|^2] \bigg\},
\label{elasticity}
\end{eqnarray}
where $c_\perp = \phi_0^2/(32\pi^3\ell^4)$, $c_y = 2\kappa_y/\ell^4$,
and $\bbox{\nabla}_\perp = (\partial_x,\partial_z)$.  The higher order
elastic correction $c_4 = \phi_0^4/(512 \pi^6\ell^4)$ may be important
experimentally due to the extremely small value of $c_y$ in the
strongly layered limit.  We observe that with screening the vortex lattice
survives down to two dimensions.

Finally, we note that in an isotropic system the low-energy degrees of
freedom are also a discrete subset $k_n$ of the LLL modes.
Even without layering, rotational invariance in the plane
perpendicular to field and the LLL physics unambiguously determine
$k_n=n\sqrt{2\pi}/l$.  One may speculate that fluctuations within such
a discrete set of modes could support a continuous freezing mechanism
within the Ginzburg-Landau theory of an isotropic superconductor as
well.

We thank Mathew Fisher and John Toner for discussions.
LR acknowledges the support by NSF DMR94-16926, through the Science
and Technology Center for Superconductivity. We are
grateful to The Institute for Theoretical Physics at UCSB, where part of
this work was done, for their support under NSF ~PHY94-07194.

\end{multicols}{}
\end{document}